\documentclass[aps,pra]{revtex4}
\usepackage{amsfonts}
\usepackage{amsmath}
\usepackage{amssymb}
\usepackage{epsfig}
\usepackage{bm}

\def\be{\begin{equation}}
\def\ee{\end{equation}}
\def\bea{\begin{eqnarray}}
\def\eea{\end{eqnarray}}

\begin{document}

\title{Parametric generation of high frequency coherent light in negative
index materials and materials with strong anomalous dispersion}
\author{Anatoly A. Svidzinsky$^{1,2}$, Xiwen Zhang$^{1}$, Luojia Wang$^{1,3}$%
, Jizhou Wang$^{1,4}$ and Marlan O. Scully$^{1,2,3}$}
\affiliation{$^1$Texas A\&M University, College Station TX 77843; $^2$Princeton
University, Princeton NJ 08544; $^3$Baylor University, Waco, TX 76706, $^4$%
Xi'an Jiaotong University, Xi'an, China}
\date{\today }

\begin{abstract}
We demonstrate the possibility of generation of coherent radiation with
tunable frequencies higher than the frequency of the driving field $\nu _{d}$
in a nonlinear medium utilizing the difference combination resonance that
occurs when $\nu _{d}$ matches the difference of the frequencies of the two
generated fields $\omega _{1}$ and $\omega _{2}$. We find that such a
resonance can appear in materials which have opposite signs of refractive
index at $\omega _{1}$ and $\omega _{2}$. It can also occur in positive
refractive index materials with strong anomalous dispersion if at one of the
generated frequencies the group and phase velocities are opposite to each
other. We show that the light amplification mechanism is equivalent to a
combination resonance in a system of two coupled parametric oscillators with
the opposite sign of masses. Such a mechanism holds promise for a new kind
of light source that emits coherent radiation of tunable wavelengths by an
optical parametric amplification process with the frequency higher than $\nu
_{d}$.
\end{abstract}

\author{}
\maketitle


\section{Introduction}

The invention of the laser in 1960's facilitated exploration of various
nonlinear phenomena in optics, such as harmonic generation~\citep{Franken61}%
, multi photon absorption~\citep{Kaiser61}, wave mixing~\citep{Maker65},
optical Kerr effect ~\citep{Maker64}, stimulated Raman scattering~%
\citep{Long02} and so on. Today nonlinear optics plays an important role in
the developing of new light sources~\citep{Garmire13}, nonlinear
spectroscopy~\citep{Mukamel95}, multiphoton microscopy~\citep{Denk90},
all-optical signal processing~\citep{Zalevsky07}, quantum information~%
\citep{DellAnno06} and many other applications~\citep{Boyd08}.

Optical parametric oscillation/amplification (OPO/OPA)~%
\citep{Kingston62,Kroll62,Akhmanov63, Giordmaine65} which transfers energy
from the driving field to the signal and idler waves is an example of
parametric processes caused by the nonlinear light-matter interaction. OPA
led to development of tunable radiation sources throughout the infrared,
visible, and ultraviolet spectral regions~\citep{Bosenberg89}. Such sources
yield amplification of coherent radiation at lower-frequencies $\omega _{1}$
and $\omega _{2}$ by pumping nonlinear optical crystals with intense laser
light at the sum frequency $\nu _{d}$%
\begin{equation}
\nu _{d}=\omega _{1}+\omega _{2}.  \label{s1}
\end{equation}%
This process is known as the sum combination resonance. It has been
discovered in electronic circuits in 1950's \citep{Heffner58,Tien58} and
later on found applications in optics 
\citep{Kingston62,Kroll62,Akhmanov63,
Giordmaine65}. Similar resonance occurs in the free electron laser in which
the low frequency plasma and laser waves are excited by the sum frequency
undulator wave in the electron rest frame \citep{Marshall85}. It also
appears when electromagnetic waves interact with optical phonons in
dielectric crystals \cite{Shal11,Popo12,Popo14}.

In the literature, another type of combination resonance, namely the
difference combination resonance, has been discussed in mechanical systems
since the 1960's (see, e.g., \cite{Hsu63,Nayf77,Nayf00,Vioque10}). It can
occur when the driving frequency matches the frequency difference between
two normal frequencies 
\begin{equation}
\nu _{d}=\omega _{2}-\omega _{1}  \label{d1}
\end{equation}%
which yields amplification of $\omega _{1}$ and $\omega _{2}$. Combination
resonances affect dynamic stability of structures and appear in various
systems having multiple degrees of freedom.

Recently, the difference combination resonance has been experimentally
demonstrated in an electronic circuit in the radio frequency range and
proposed as a mechanism of generation of high frequency coherent radiation
(e.g. XUV or X-rays) utilizing a low frequency (e.g., Infrared) driving
field \citep{Svidzinsky13}. Such a parametric amplifier, which operates by
Quantum Amplification by Superradiant Emission of Radiation, is called the
QASER for short. In contrast to OPOs, QASER generates light at higher
frequencies than $\nu _{d}$.

Here we study the possibility of generation of high frequency coherent
radiation using the difference combination resonance in nonlinear
metamaterials. Metamaterials are artificial inhomogeneous structures
composed of periodic subwavelength, polarizable elements. They can give rise
to negative refractive index, near-zero permittivity or permeability of the
macroscopic response in the microwave, terahertz and even optical regions~%
\citep{Dolling06,Xiao09,Edwards08} and have led to a variety of fascinating
phenomena such as a prefect lens and invisible cloaks~%
\citep{Smith04,Ramakrishma05,Liu11}. Inclusion of nonlinear elements into
metamaterials offers unique advantages to enhance nonlinearities by
local-field amplification~\citep{Pendry99,Rose11pra} and to provide
particular constructive configurations under the phase-matching condition
for highly efficient nonlinear processes %
\citep{Rose11,Popov06,Rose11prl,Suchowski13}. Anomalous transmission
properties of zero-permittivity channels can be used to boost Kerr
nonlinearities \citep{Argy12}. Effect of nonlinearity in combination with
the unique linear properties has been recently investigated in connection
with new frequencies generation~\citep{Rose11prl,Huang11,Vincenti11},
solitary wave propagation~\citep{Lazarides06,Liu07}, self tuning and
hysteresis transition~\citep{Zharov03,Powell07} and parametric amplification~%
\citep{Popov06ol,Litchinitser09,Poutrina10}.

Study of parametric amplification in such materials has demonstrated that if
for one of the waves the refractive index is negative (directions of the
wave vector and the Poynting vector are opposite) the phase matching leads
to a possibility of loss compensation and distributed-feedback parametric
oscillation with no cavity~\citep{Popov06,Popov06ol,Litchinitser09}. In such
system the backward-propagating wave provides an automatic feedback
mechanism and yields dynamics analogous to the quasi-phase matched
mirror-less OPO~\citep{Ding96}.

\begin{figure}[h]
\begin{center}
\epsfig{figure=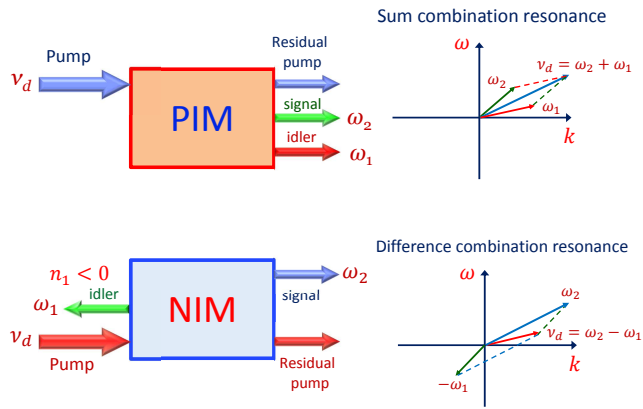, angle=270, width=9.0 cm}
\end{center}
\caption{Generation of lower frequencies $\protect\omega _{1}$ and $\protect%
\omega _{2}$ in a positive index nonlinear material (PIM) by the sum
combination resonance (top). Generation of light at higher frequencies in a
negative refractive index nonlinear material (NIM) due to the difference
combination resonance (bottom). One can physically interpret the latter
process as annihilation of the driving field photon with energy $\hbar 
\protect\nu _{d}$ and creation of two photons - one with positive energy $%
\hbar \protect\omega _{2}$ and one with negative energy $-\hbar \protect%
\omega _{1}$.}
\label{MFig1}
\end{figure}

Here we investigate parametric amplification in metamaterials caused by the
difference combination resonance. We find that when for the idler (frequency 
$\omega _{1}$) and the signal (frequency $\omega _{2}$) waves the refractive
index has opposite signs they both can be amplified by a strong driving
field with frequency $\nu _{d}=\omega _{2}-\omega _{1}$ (see Fig. \ref{MFig1}%
). We show that the light amplification mechanism in negative index
materials is equivalent to a combination resonance in a system of two
coupled parametric oscillators with the opposite sign of masses.

We also find that the difference combination resonance can be achieved in
materials with positive refractive index but strong anomalous dispersion.
Namely, it can occur if at one of the generated frequencies $\omega _{1}$ or 
$\omega _{2}$ the group velocity is opposite to the phase velocity. This
condition is somewhat easier to realize than the negative refractive index.
Photonic crystals \cite{Stei93} and fiber Bragg gratings \cite{Long03} are
examples of materials with strong anomalous dispersion. Electromagnetically
induced transparency (EIT) is a popular technique to decrease absorption
near a resonance in the region of strong dispersion. Using the EIT
technique, the group velocity of light as low as meters per second or even
completely stopped light was demonstrated in ultracold gases \cite%
{Hau99,Liu01}, hot gases \cite{Kash99,Phil01} and in a very cold solid \cite%
{Turu02}. Negative group velocity can be also obtained in the Kerr medium,
which possesses a large nonlinear refractive index and long relaxation time,
such as Cr-doped alexandrite, ruby and GdAlO$_{3}$ \cite{Yang05}.

Parametric amplification process due to the difference combination resonance
holds promise for a new kind of higher frequency convertor with tunable
wavelength that makes use of metamaterials with negative refractive index or
strongly dispersive media.

\section{Parametric generation of high frequency coherent radiation in
nonlinear materials}

\subsection{Combination resonances for two coupled parametric oscillators}

Generation of high frequencies by a parametric resonance in systems with
several normal modes can be understood in a simple example of two classical
oscillators with masses $m_{1}$ and $m_{2}$ and (positive) frequencies $%
\omega _{1}$ and $\omega _{2}$ that are weakly coupled with a periodically
varying strength. The Hamiltonian of such a system reads%
\begin{equation}
H=\frac{p_{1}^{2}}{2m_{1}}+\frac{1}{2}m_{1}\omega _{1}^{2}x_{1}^{2}+\frac{%
p_{2}^{2}}{2m_{2}}+\frac{1}{2}m_{2}\omega _{2}^{2}x_{2}^{2}+F(t)x_{1}x_{2}.
\end{equation}%
Using the Hamilton equations of motion and rescaling the coordinate $%
x_{1}\rightarrow \sqrt{|m_{2}/m_{1}|}x_{1}$ we obtain%
\begin{equation}
\ddot{x}_{1}+\omega _{1}^{2}x_{1}+f_{12}(t)x_{2}=0,  \label{x1}
\end{equation}%
\begin{equation}
\ddot{x}_{2}+\omega _{2}^{2}x_{2}+f_{21}(t)x_{1}=0,  \label{x2}
\end{equation}%
where 
\begin{equation}
f_{12}(t)=\frac{F(t)}{\sqrt{|m_{1}m_{2}|}}\text{sign}(m_{2}),
\end{equation}%
and%
\begin{equation}
f_{21}=\text{sign}(m_{1}m_{2})f_{12}  \label{xx}
\end{equation}%
describe coupling between oscillators. For harmonic modulation of the
coupling strength with frequency $\nu _{d}$ we write 
\begin{equation}
f_{12}=\delta \cos (\nu _{d}t),  \label{x3}
\end{equation}%
where $\delta $ is the modulation amplitude. According to Eq. (\ref{xx}),
the coupling between oscillators is symmetric ($f_{21}=f_{12}$) if $%
m_{1}m_{2}>0$ and antisymmetric ($f_{21}=-f_{12}$) if the masses $m_{1}$ and 
$m_{2}$ have opposite signs \cite{S14}.

Next we assume that modulation frequency obeys the condition of the
combination parametric resonance 
\begin{equation}
\nu _{d}=\omega _{2}\pm \omega _{1}.  \label{r1}
\end{equation}%
Making a slowly varying amplitude approximation, namely writing 
\begin{equation}
x_{1}(t)=A_{1}(t)\cos (\omega _{1}t),
\end{equation}%
\begin{equation}
x_{2}(t)=A_{2}(t)\sin (\omega _{2}t),
\end{equation}%
where $A_{1}(t)$ and $A_{2}(t)$ are slowly varying envelope functions on a
time scale $1/|\omega _{2}-\omega _{1}|$ and keeping only the resonant
terms, Eqs. (\ref{x1}) and (\ref{x2}) reduce to%
\begin{equation}
\dot{A}_{1}\pm \frac{\delta }{4\omega _{1}}A_{2}=0,
\end{equation}%
\begin{equation}
\dot{A}_{2}+\text{sign}(m_{1}m_{2})\frac{\delta }{4\omega _{2}}A_{1}=0,
\end{equation}%
which yield the following equation for $A_{1}(t)$%
\begin{equation}
\ddot{A}_{1}\mp \text{sign}(m_{1}m_{2})\frac{\delta ^{2}}{16\omega
_{1}\omega _{2}}A_{1}=0.  \label{x4}
\end{equation}%
Solutions for $A_{1,2}$ exponentially grow with time, $A_{1,2}\propto e^{Gt}$%
, provided $\pm m_{1}m_{2}>0$. Namely, for symmetric coupling ($m_{1}m_{2}>0$%
), the solution grows if $\nu _{d}=\omega _{2}+\omega _{1}$ (the sum
combination resonance). For antisymmetric coupling ($m_{1}m_{2}<0$) there is
exponential grow provided $\nu _{d}=\omega _{2}-\omega _{1}$ (the difference
combination resonance). For both resonances the gain per unit time $G$ is
given by%
\begin{equation}
G=\frac{\delta }{4\sqrt{\omega _{1}\omega _{2}}}.  \label{x5}
\end{equation}

The sum combination resonance is used for generation of frequencies lower
than $\nu _{d}$. This is the case for optical parametric oscillators that
convert an input laser wave with frequency $\nu _{d}$ into two output waves
of lower frequency ($\omega _{1}$ and $\omega _{2}$) by means of the second
order nonlinear optical interaction. On the other hand, the difference
combination resonance can be used for the generation of higher frequencies.
This has been recently demonstrated in nonreciprocally coupled electronic
circuits in the radio frequency domain \citep{Svidzinsky13}.

\begin{figure}[h]
\begin{center}
\epsfig{figure=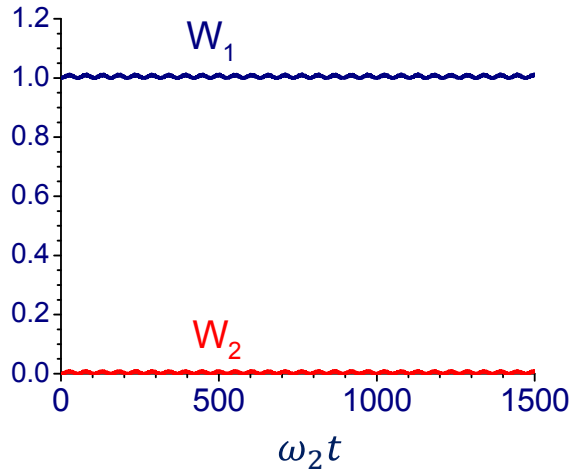, angle=270, width=9.0 cm}
\end{center}
\caption{Energy $W_{1}$ of the first and of the second $W_{2}$ oscillator
[in units of $W_{1}(0)$] as a function of time for asymmetric coupling $%
f_{12}=-f_{21}=\protect\delta \cos (\protect\nu _{d}t)$ and off-resonance
modulation $\protect\nu _{d}=0.35\protect\omega _{2}$ obtained by numerical
solution of Eqs. (\protect\ref{x1}) and (\protect\ref{x2}) with $\protect%
\omega _{1}=0.77\protect\omega _{2}$ and $\protect\delta =0.02\protect\omega%
^2 _{2}$. Initially the first oscillator is excited while the second
oscillator is at rest.}
\label{MFig2}
\end{figure}

\begin{figure}[h]
\begin{center}
\epsfig{figure=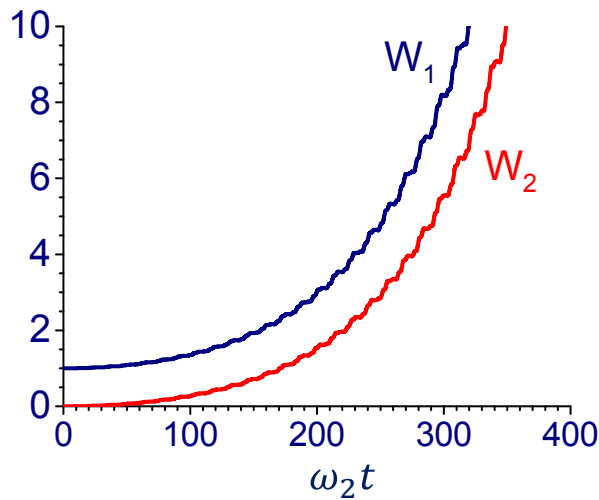, angle=270, width=9.0 cm}
\end{center}
\caption{The same as in Fig. \protect\ref{MFig2} but for the resonant
modulation $\protect\nu _{d}=\protect\omega_2-\protect\omega_1=0.23\protect%
\omega _{2}$.}
\label{MFig3}
\end{figure}

\begin{figure}[h]
\begin{center}
\epsfig{figure=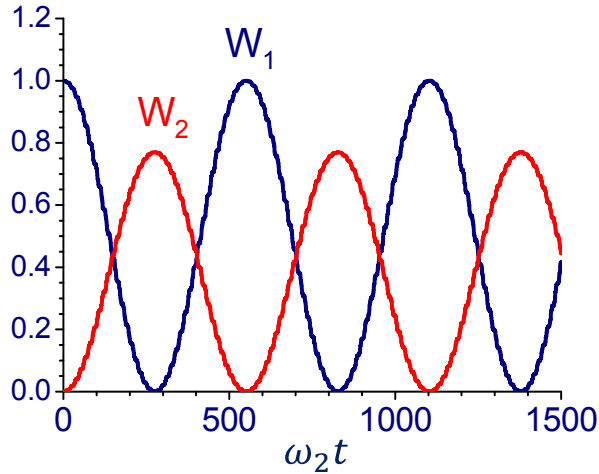, angle=270, width=9.0 cm}
\end{center}
\caption{The same as in Fig. \protect\ref{MFig3} but for symmetric coupling $%
f_{12}=f_{21}=\protect\delta \cos (\protect\nu _{d}t)$.}
\label{MFig4}
\end{figure}

\begin{figure}[h]
\begin{center}
\epsfig{figure=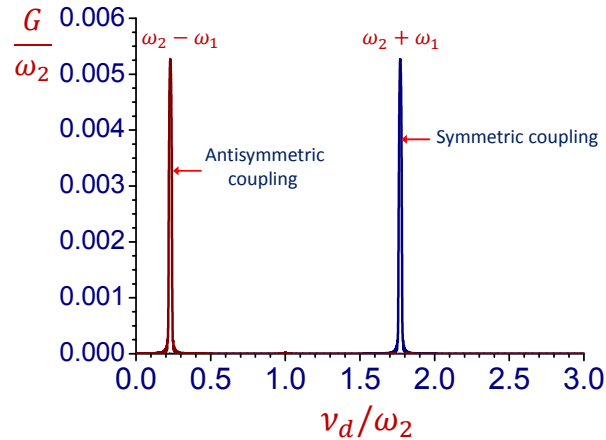, angle=270, width=9.0 cm}
\end{center}
\caption{Gain $G$ as a function of the modulation frequency $\protect\nu %
_{d} $ for symmetric ($f_{12}=f_{21}$) and antisymmetric ($f_{12}=-f_{21}$)
coupling between oscillators obtained from numerical solution of Eqs. (%
\protect\ref{x1}), (\protect\ref{x2}) and (\protect\ref{x3}) with $\protect%
\omega _{1}=0.77\protect\omega _{2}$ and $\protect\delta =0.02\protect\omega %
_{2}^{2}$. For symmetric coupling the system has the sum combination
resonance at $\protect\nu _{d}=\protect\omega _{2}+\protect\omega _{1}$,
while for the antisymmetric coupling there is the difference combination
resonance at $\protect\nu _{d}=\protect\omega _{2}-\protect\omega _{1}$.}
\label{MFig5}
\end{figure}

As an illustration of combination resonances, in Figs. \ref{MFig2}-\ref%
{MFig4} we plot energy of the first $W_{1}$ and the second $W_{2}$
oscillator as a function of time for off-resonance (Fig. \ref{MFig2}) and
on-resonance (Figs. \ref{MFig3} and \ref{MFig4}) modulation obtained by
numerical solution of Eqs. (\ref{x1}) and (\ref{x2}). In simulations we take 
$\omega _{1}=0.77\omega _{2}$ and $\delta =0.02\omega _{2}^{2}$. Initially
the first oscillator is excited while the second oscillator is at rest. If $%
\nu _{d}$ does not satisfy the resonance condition (\ref{r1}) the
oscillators do not exchange their energy, as shown in Fig. \ref{MFig2}.

However, if, e.g., the condition of the difference combination resonance is
satisfied, $\nu _{d}=\omega _{2}-\omega _{1}$, the oscillator's energies
exponentially grow for antisymmetric coupling (see Fig. \ref{MFig3}). This
is the essence of the QASER amplification mechanism \citep{Svidzinsky13}. If
the coupling is symmetric, the difference combination resonance yields an
efficient energy exchange between two oscillators without exponential grow,
as illustrated in Fig. \ref{MFig4}. This can be used in various
applications, e.g., for fast control of propagation of $\gamma $-rays
through crystals \citep{Zhang13}.

In Fig. \ref{MFig5} we plot the gain $G$ as a function of the modulation
frequency $\nu _{d}$ for symmetric ($f_{12}=f_{21}$) and antisymmetric ($%
f_{12}=-f_{21}$) coupling between oscillators. For symmetric coupling the
system undergoes the sum combination resonance at $\nu _{d}=\omega
_{2}+\omega _{1}$, but no difference combination resonance. On the other
hand, for the antisymmetric coupling between oscillators there is the
difference combination resonance at $\nu _{d}=\omega _{2}-\omega _{1}$, but
no resonance at $\nu _{d}=\omega _{2}+\omega _{1}$.

\subsection{Combination resonances in magneto-optical crystals}

The possibility of a difference combination resonance in materials with
negative refractive index can be demonstrated in the simple example of
magneto-optical effect. In a magneto-optic material the presence of an
externally applied magnetic field $H_{0}(t)$ causes a change in the
permittivity tensor $\varepsilon _{ik}$. The tensor becomes anisotropic with
complex off-diagonal components. The off-diagonal components of $\varepsilon
_{ik}$ describe coupling between two perpendicular polarizations of an
electromagnetic wave. Such coupling can be periodically modulated if the
externally applied magnetic field is a harmonic function of time $%
H_{0}\propto \cos (\nu _{d}t)$.

Here we show that the problem of propagation of electromagnetic waves
through the gyromagnetic medium can be reduced to the two coupled parametric
oscillators discussed in the previous section. Evolution of the
electromagnetic field in the medium is described by Maxwell's equations%
\begin{equation}
\text{curl}\mathbf{E}=-\frac{\partial \mathbf{B}}{\partial t},\qquad \text{%
curl}\mathbf{H}=\frac{\partial \mathbf{D}}{\partial t}.  \label{me}
\end{equation}%
We assume that the wave propagates along the $z$-axis and the
electromagnetic field of the wave has only $x$ and $y$ components which
depend on $t$ and $z$. Then the Maxwell's equations (\ref{me}) reduce to%
\begin{equation}
\frac{\partial E_{y}}{\partial z}=\frac{\partial B_{x}}{\partial t},\qquad 
\frac{\partial E_{x}}{\partial z}=-\frac{\partial B_{y}}{\partial t},
\label{me1}
\end{equation}%
\begin{equation}
\frac{\partial H_{y}}{\partial z}=-\frac{\partial D_{x}}{\partial t},\qquad 
\frac{\partial H_{x}}{\partial z}=\frac{\partial D_{y}}{\partial t}.
\label{me2}
\end{equation}%
If the absorption losses can be neglected, the permittivity tensor is a
Hermitian matrix \citep{Tsao93} and we can write%
\begin{equation}
\left( 
\begin{array}{c}
E_{x} \\ 
E_{y}%
\end{array}%
\right) =\left( 
\begin{array}{cc}
\tilde{\varepsilon}_{x} & ig(t) \\ 
-ig(t) & \tilde{\varepsilon}_{y}%
\end{array}%
\right) \left( 
\begin{array}{c}
D_{x} \\ 
D_{y}%
\end{array}%
\right) ,  \label{me3}
\end{equation}%
\begin{equation}
\left( 
\begin{array}{c}
H_{x} \\ 
H_{y}%
\end{array}%
\right) =\left( 
\begin{array}{cc}
\tilde{\mu}_{x} & 0 \\ 
0 & \tilde{\mu}_{y}%
\end{array}%
\right) \left( 
\begin{array}{c}
B_{x} \\ 
B_{y}%
\end{array}%
\right) ,  \label{me4}
\end{equation}%
where the off-diagonal component $g(t),$ to first order, is proportional to
the applied external magnetic field $H_{0}(t)$ that lies along the $z$-axis %
\citep{LL8}. Periodic variation of $H_{0}(t)$ with time yields periodic
modulation of $g$, $g(t)=g_{0}\cos (\nu _{d}t)$. Taking time derivative from
Eqs. (\ref{me2}) and using Eqs. (\ref{me1}), (\ref{me3}) and (\ref{me4}) we
obtain 
\begin{equation}
\frac{\partial ^{2}D_{x}}{\partial t^{2}}-\tilde{\mu}_{y}\tilde{\varepsilon}%
_{x}\frac{\partial ^{2}D_{x}}{\partial z^{2}}-ig(t)\tilde{\mu}_{y}\frac{%
\partial ^{2}D_{y}}{\partial z^{2}}=0,
\end{equation}%
\begin{equation}
\frac{\partial ^{2}D_{y}}{\partial t^{2}}-\tilde{\mu}_{x}\tilde{\varepsilon}%
_{y}\frac{\partial ^{2}D_{y}}{\partial z^{2}}+ig(t)\tilde{\mu}_{x}\frac{%
\partial ^{2}D_{x}}{\partial z^{2}}=0.
\end{equation}%
Change of functions%
\begin{equation}
D_{x}=i\sqrt{|\tilde{\mu}_{y}|}\tilde{D}_{x},\qquad D_{y}=\sqrt{|\tilde{\mu}%
_{x}|}\tilde{D}_{y}
\end{equation}%
yields%
\begin{equation}
\frac{\partial ^{2}\tilde{D}_{x}}{\partial t^{2}}-\tilde{\mu}_{y}\tilde{%
\varepsilon}_{x}\frac{\partial ^{2}\tilde{D}_{x}}{\partial z^{2}}-\text{sign}%
(\tilde{\mu}_{y})\sqrt{|\tilde{\mu}_{x}\tilde{\mu}_{y}|}g(t)\frac{\partial
^{2}\tilde{D}_{y}}{\partial z^{2}}=0,
\end{equation}%
\begin{equation}
\frac{\partial ^{2}\tilde{D}_{y}}{\partial t^{2}}-\tilde{\mu}_{x}\tilde{%
\varepsilon}_{y}\frac{\partial ^{2}\tilde{D}_{y}}{\partial z^{2}}-\text{sign}%
(\tilde{\mu}_{x})\sqrt{|\tilde{\mu}_{x}\tilde{\mu}_{y}|}g(t)\frac{\partial
^{2}\tilde{D}_{x}}{\partial z^{2}}=0.
\end{equation}%
One can look for a solution of these equations in the form

\begin{equation}
\tilde{D}_{x,y}(t,z)=\tilde{D}_{x,y}(t)e^{ikz}
\end{equation}%
which results in the equations of two coupled harmonic oscillators with time
dependent coupling%
\begin{equation}
\frac{\partial ^{2}\tilde{D}_{x}}{\partial t^{2}}+\omega _{1}^{2}\tilde{D}%
_{x}+f_{12}(t)\tilde{D}_{y}=0,  \label{x6}
\end{equation}%
\begin{equation}
\frac{\partial ^{2}\tilde{D}_{y}}{\partial t^{2}}+\omega _{2}^{2}\tilde{D}%
_{y}+f_{21}(t)\tilde{D}_{x}=0,  \label{x7}
\end{equation}%
where%
\begin{equation}
\omega _{1}^{2}=\tilde{\mu}_{y}\tilde{\varepsilon}_{x}k^{2},\qquad \omega
_{2}^{2}=\tilde{\mu}_{x}\tilde{\varepsilon}_{y}k^{2},  \label{x8}
\end{equation}%
\begin{equation}
f_{12}(t)=\sqrt{|\tilde{\mu}_{x}\tilde{\mu}_{y}|}k^{2}\text{sign}(\tilde{\mu}%
_{y})g(t)\propto \cos (\nu _{d}t),  \label{x9}
\end{equation}%
and%
\begin{equation}
f_{21}(t)=\text{sign}(\tilde{\mu}_{x}\tilde{\mu}_{y})f_{12}(t).  \label{x10}
\end{equation}

Eqs. (\ref{x6}) and (\ref{x7}) are identical to the equations of the coupled
parametric oscillators (\ref{x1}) and (\ref{x2}) investigated in the
previous section and, hence, they display the same combination resonances
when the modulation frequency satisfies the condition $\nu _{d}=\omega
_{2}\pm \omega _{1}$. Namely, if $\tilde{\mu}_{x}\tilde{\mu}_{y}>0$ the
coupling between the two field polarizations $x$ and $y$ is symmetric and
there is the sum combination resonance at $\nu _{d}=\omega _{2}+\omega _{1}$
which yields generation of lower frequencies $\omega _{1}$ and $\omega _{2}$%
. This is the case for the positive index materials and is an operation
mechanism of the optical parametric amplifiers.

The coupling between the two field polarizations is antisymmetric if $\tilde{%
\mu}_{x}$ and $\tilde{\mu}_{y}$ have opposite signs. To make frequencies in
Eq. (\ref{x8}) real the corresponding components of the permittivity tensor
also should have opposite signs. Namely, if $\tilde{\mu}_{x}<0$ and $\tilde{%
\mu}_{y}>0$ then we must have $\tilde{\varepsilon}_{y}<0$ and $\tilde{%
\varepsilon}_{x}>0$. Under these conditions there is the difference
combination resonance at $\nu _{d}=\omega _{2}-\omega _{1}$ which yields
generation of higher frequencies $\omega _{1}$ and $\omega _{2}$. This case
is similar to the two coupled oscillators with the opposite sign of masses
discussed in the previous section.

According to Eq. (\ref{x8}), generated waves at frequencies $\omega _{1}$
and $\omega _{2}$ have the same wave vector $\mathbf{k}$, but orthogonal
polarizations. The values of $\omega _{1}$ and $\omega _{2}$ differ due to
the anisotropy of the medium ($\tilde{\mu}_{y}\tilde{\varepsilon}_{x}\neq 
\tilde{\mu}_{x}\tilde{\varepsilon}_{y}$). Using combination resonance one
can also generate waves with different wave vectors $\mathbf{k}_{1}$ and $%
\mathbf{k}_{2}$ if the driving field wave vector $\mathbf{k}_{d}$
compensates for the momentum mismatch $\mathbf{k}_{d}=\mathbf{k}_{2}\pm 
\mathbf{k}_{1}$.

Moreover, if we take into account frequency dispersion the expression for
the gain $G$ is modified. Straightforward calculations similar to those we
present in the next section yield

\begin{equation}
G^{2}=\pm \frac{\omega _{1}\omega _{2}g_{0}^{2}}{16\tilde{\varepsilon}%
_{x}(\omega _{1})\tilde{\varepsilon}_{y}(\omega _{2})}\frac{V_{g1}V_{g2}}{%
V_{p1}V_{p2}},  \label{g1}
\end{equation}%
where $V_{g1,2}$ and $V_{p1,2}$ are the group and phase velocities of the
waves with frequencies $\omega _{1,2}$. Eq. (\ref{g1}) shows that the
difference combination resonance can also occur in materials with positive
refractive index if at one of the generated frequencies the group and the
phase velocities are opposite to each other. This requires medium with
strong anomalous dispersion. We discuss these questions in details in the
next section in which we consider combination resonances in nonlinear
dispersive medium assuming that waves have the same linear polarization.

\subsection{Combination resonances in nonlinear dispersive medium}

Using nonlinear crystals the optical parametric oscillators generate light
at frequencies $\omega _{1}$ and $\omega _{2}$ satisfying the condition of
the sum combination resonance $\nu _{d}=\omega _{1}+\omega _{2}$, where $\nu
_{d}$ is the frequency of the strong driving (pump) field $\mathbf{E}_{d}$.
Here we investigate the possibility of difference combination resonance in
such materials. In the present model we assume that medium is spatially
uniform, isotropic and polarization is a nonlinear function of the electric
field, while magnetization is a linear function of the magnetic field. We
also assume that the driving field and generated waves propagate along the $%
z $-axis and have the same polarization. Under these conditions the
Maxwell's equations reduce to%
\begin{equation}
\frac{\partial E}{\partial z}=\frac{\partial B}{\partial t},\qquad \frac{%
\partial H}{\partial z}=\frac{\partial D}{\partial t}.  \label{d1m}
\end{equation}%
We write the total electric field as%
\begin{equation}
E=E_{d}+E_{1}+E_{2},  \label{s6}
\end{equation}%
where the strong driving field is assumed to be fixed%
\begin{equation}
E_{d}=A_{d}\cos \left( \nu _{d}t-k_{d}z\right)
\end{equation}%
and $E_{1,2}\ll $ $E_{d}$. In nonlinear isotropic medium the total
displacement field in the absence of dispersion can be written as 
\begin{equation*}
D(t,z)=\varepsilon \varepsilon _{0}E+\chi \varepsilon _{0}E^{2}\approx
\varepsilon \varepsilon _{0}\left( E_{d}+E_{1}+E_{2}\right) +\chi
\varepsilon _{0}E_{d}^{2}
\end{equation*}%
\begin{equation*}
+2\chi \varepsilon _{0}E_{1}A_{d}\cos \left( \nu _{d}t-k_{d}z\right) +2\chi
\varepsilon _{0}E_{2}A_{d}\cos \left( \nu _{d}t-k_{d}z\right) ,
\end{equation*}%
where we omitted small terms of the second order in $E_{1,2}$. $\varepsilon $
is the linear permittivity of the material and $\chi $ is the nonlinear
susceptibility. If there is frequency dispersion it is convenient to write
the Maxwell's equations (\ref{d1m}) and the constitutive relations in terms
of the field Fourier components, namely%
\begin{equation}
kE(\omega ,k)=-\omega B(\omega ,k),\qquad kH(\omega ,k)=-\omega D(\omega ,k),
\label{d2}
\end{equation}%
\begin{equation}
B(\omega ,k)=\mu (\omega )\mu _{0}H(\omega ,k),  \label{d3}
\end{equation}%
\begin{equation*}
D(\omega ,k)=\varepsilon (\omega )\varepsilon _{0}E(\omega ,k)+
\end{equation*}%
\begin{equation}
\chi \varepsilon _{0}A_{d}\left[ E_{1}(\omega -\nu
_{d},k-k_{d})+E_{1}(\omega +\nu _{d},k+k_{d})+E_{2}(\omega -\nu
_{d},k-k_{d})+E_{2}(\omega +\nu _{d},k+k_{d})\right] ,  \label{d4}
\end{equation}%
where we disregarded irrelevant term proportional to $E_{d}^{2}$, took into
account that Fourier transform of $e^{-i\nu _{d}t+ik_{d}z}f(t,z)$ gives $%
f(\omega -\nu _{d},k-k_{d})$ and for simplicity assumed that the nonlinear
susceptibility $\chi $ has no dispersion. In our considerations the
wavenumber $k$ is real while frequency $\omega $ has small imaginary part $G$
which determines the wave amplification or attenuation per unit time. Plug
Eqs. (\ref{d3}) and (\ref{d4}) into Eq. (\ref{d2}) yields the following
equation for $E_{1}$ and $E_{2}$%
\begin{equation*}
\left[ c^{2}k^{2}-\omega ^{2}\varepsilon (\omega )\mu (\omega )\right]
\left( E_{1}(\omega ,k)+E_{2}(\omega ,k)\right)
\end{equation*}%
\begin{equation}
-\omega ^{2}\chi \mu (\omega )A_{d}\left[ E_{1}(\omega -\nu
_{d},k-k_{d})+E_{1}(\omega +\nu _{d},k+k_{d})+E_{2}(\omega -\nu
_{d},k-k_{d})+E_{2}(\omega +\nu _{d},k+k_{d})\right] =0.  \label{d5}
\end{equation}

Next we assume that 
\begin{equation}
\nu _{d}=\omega _{2}\pm \omega _{1},  \label{s1a}
\end{equation}%
\begin{equation}
k_{d}=k_{2}\pm k_{1},  \label{s2a}
\end{equation}%
where $k_{1,2}$ and $\omega _{1,2}$ obey the relations%
\begin{equation}
c^{2}k_{1}^{2}=\omega _{1}^{2}\varepsilon (\omega _{1})\mu (\omega _{1}),
\label{s3}
\end{equation}%
\begin{equation}
c^{2}k_{2}^{2}=\omega _{2}^{2}\varepsilon (\omega _{2})\mu (\omega _{2}).
\label{s4}
\end{equation}%
Since the driving field propagates through the same medium its frequency $%
\nu _{d}$ and the wavenumber $k_{d}$ satisfy the similar equation%
\begin{equation}
c^{2}k_{d}^{2}=\nu _{d}^{2}\varepsilon (\nu _{d})\mu (\nu _{d}).  \label{s5}
\end{equation}%
Eqs. (\ref{s1a}) and (\ref{s2a}) are consistent with Eqs. (\ref{s3})-(\ref%
{s5}) provided%
\begin{equation}
\omega _{2}\left( 1-\sqrt{\frac{\varepsilon _{2}\mu _{2}}{\varepsilon
_{d}\mu _{d}}}\right) =\mp \omega _{1}\left( 1-\sqrt{\frac{\varepsilon
_{1}\mu _{1}}{\varepsilon _{d}\mu _{d}}}\right) ,  \label{c1}
\end{equation}%
where $\varepsilon _{1,2}$ and $\varepsilon _{d}$ stands for $\varepsilon
(\omega _{1,2})$ and $\varepsilon (\nu _{d})$, etc.

We look for a solution of Eq. (\ref{d5}) in the form%
\begin{equation}
E_{1}(\omega ,k)\approx E_{1}\delta (\omega \pm \omega _{1}-iG)\delta (k\pm
k_{1}),\qquad E_{2}(\omega ,k)\approx E_{2}\delta (\omega -\omega
_{2}-iG)\delta (k-k_{2}),  \label{d11}
\end{equation}%
where $G$ is the gain per unit time. Keeping the resonant terms we obtain 
\begin{equation}
\left[ \left( c^{2}k^{2}-\omega ^{2}\varepsilon (\omega )\mu (\omega
)\right) E_{1}-\omega ^{2}\chi \mu (\omega )A_{d}E_{2}\right] \delta (\omega
\pm \omega _{1}-iG)\delta (k\pm k_{1})=0,  \label{h1}
\end{equation}%
\begin{equation}
\left[ \left( c^{2}k^{2}-\omega ^{2}\varepsilon (\omega )\mu (\omega
)\right) E_{2}-\omega ^{2}\chi \mu (\omega )A_{d}E_{1}\right] \delta (\omega
-\omega _{2}-iG)\delta (k-k_{2})=0.  \label{h2}
\end{equation}%
The factor in front of the delta functions in Eq. (\ref{h1}) must be
calculated at $k=\mp k_{1}$ and $\omega =\mp \omega _{1}+iG$, while the
factor in Eq. (\ref{h2}) should be estimated at $k=k_{2}$ and $\omega
=\omega _{2}+iG$. Making a Taylor expansion one can write%
\begin{equation}
\left. \left( c^{2}k^{2}-\omega ^{2}\varepsilon (\omega )\mu (\omega
)\right) \right\vert _{k=\mp k_{1},\omega =\mp \omega _{1}+iG}\approx \pm iG%
\frac{\partial }{\partial \omega _{1}}\left[ \omega _{1}^{2}\varepsilon
(\omega _{1})\mu (\omega _{1})\right] =\pm iG\frac{2c\omega _{1}n_{1}}{V_{g1}%
},  \label{g2}
\end{equation}%
\begin{equation}
\left. \left( c^{2}k^{2}-\omega ^{2}\varepsilon (\omega )\mu (\omega
)\right) \right\vert _{k=k_{2},\omega =\omega _{2}+iG}\approx -iG\frac{%
\partial }{\partial \omega _{2}}\left[ \omega _{2}^{2}\varepsilon (\omega
_{2})\mu (\omega _{2})\right] =-iG\frac{2c\omega _{2}n_{2}}{V_{g2}},
\label{g3}
\end{equation}%
where $|n(\omega )|=\sqrt{\varepsilon (\omega )\mu (\omega )}$ is the
refractive index ($n$ is positive for $\varepsilon $, $\mu >0$ and negative
for $\varepsilon $, $\mu <0$), 
\begin{equation}
V_{g}(\omega )=\frac{\partial \omega }{\partial k}=\frac{2c\omega n(\omega )%
}{\frac{\partial \left( \omega ^{2}\varepsilon \mu \right) }{\partial \omega 
}}
\end{equation}%
is the group velocity of the wave at frequency $\omega $ and $c=1/\sqrt{%
\varepsilon _{0}\mu _{0}}$ is the speed of light in vacuum. For brevity we
introduced notations $n_{1,2}=n(\omega _{1,2})$ and $V_{g1,2}=V_{g}(\omega
_{1,2})$. Taking into account Eqs. (\ref{g2}) and (\ref{g3}), and keeping
the leading terms, Eqs. (\ref{h1}) and (\ref{h2}) reduce to%
\begin{equation}
\pm iG\frac{2cn_{1}}{V_{g1}}E_{1}-\omega _{1}\chi \mu _{1}A_{d}E_{2}=0,
\label{d8}
\end{equation}%
\begin{equation}
iG\frac{2cn_{2}}{V_{g2}}E_{2}+\omega _{2}\chi \mu _{2}A_{d}E_{1}=0.
\label{d9}
\end{equation}%
Eqs. (\ref{d8}) and (\ref{d9}) give the following expression for the gain%
\begin{equation}
G^{2}=\pm \mu _{1}\mu _{2}\omega _{1}\omega _{2}\chi ^{2}A_{d}^{2}\frac{%
V_{g1}V_{p1}V_{g2}V_{p2}}{4c^{4}},  \label{s12}
\end{equation}%
where $V_{p1,2}=c/n_{1,2}$ is the phase velocity at frequency $\omega _{1,2}$%
. In the time domain solution (\ref{d11}) reads 
\begin{equation}
E_{1}(t,z)=E_{1}e^{\pm i\omega _{1}t\mp ik_{1}z+Gt},\qquad
E_{2}(t,z)=E_{2}e^{-i\omega _{2}t+ik_{2}z+Gt}.
\end{equation}%
This solution exponentially grows with time if $G$ is real (and positive).
Thus, in order to have amplification the right hand side of Eq. (\ref{s12})
must be positive. For materials with $\mu _{1,2}>0$ there is gain for the
upper sign, that is when $\nu _{d}=\omega _{2}+\omega _{1}$. Such materials
yield the sum combination resonance and are used in conventional optical
parametric amplifiers. In order to achieve light amplification at the
difference combination resonance $\nu _{d}=\omega _{2}-\omega _{1}$ (the
lower sign in Eq. (\ref{s12})) we must have%
\begin{equation}
\mu _{1}\mu _{2}<0.  \label{c0}
\end{equation}

Conditions (\ref{c0}) and (\ref{c1}) can be fulfilled, e.g., if the crystal
has a negative refractive index at the frequency $\omega _{1}$ ($\varepsilon
_{1}<0$ and $\mu _{1}<0$), but a positive refractive index at the frequency $%
\omega _{2}$ ($\varepsilon _{2}>0$ and $\mu _{2}>0$). In addition, Eq. (\ref%
{s12}) shows that the difference combination resonance can also occur in
nonlinear materials with positive refractive index if at one of the
generated frequencies $\omega _{1}$ or $\omega _{2}$ the group velocity $%
V_{g}$ is opposite to the phase velocity $V_{p}$. This condition is somewhat
easier to achieve than the negative refractive index. It requires strong
anomalous dispersion, however, material does not need to have both magnetic
and electric responses.

In our analysis we have assumed that medium is spatially uniform. An
interesting question appears in this connection. Namely, if a material is
nonuniform can this eliminate the requirement of having a negative
refractive index to achieve the difference combination resonance in medium
with weak dispersion? In Appendix A we investigate this question for
materials described by the local constitutive relations $\mathbf{D}%
=\varepsilon (\mathbf{r})\varepsilon _{0}\mathbf{E}$ and $\mathbf{B}=\mu (%
\mathbf{r})\mu _{0}\mathbf{H}$. Based on a general consideration we show
that the difference combination resonance in nonlinear weakly dispersive
materials is possible only if for one of the generated frequencies both the
magnetic permeability $\mu $ and dielectric permittivity $\varepsilon $ are
negative in some region of space.

\section{Summary}

In this paper we consider parametric generation of coherent radiation in
nonlinear medium produced by a strong driving field. In conventional
nonlinear crystals with a positive refractive index such parametric
mechanism yields generation of smaller frequencies due to the sum
combination resonance (see Eq. (\ref{s1})) which is the operating principle
of optical parametric amplifiers. Here we find that in left-handed materials
the difference combination resonance described by Eq. (\ref{d1}) can occur
if for one of the generated frequencies the refractive index is negative.
Such a resonance yields excitation of frequencies higher than $\nu _{d}$.

We show that this mechanism is similar to a combination resonance in a
system of two coupled parametric oscillators with periodically modulated
coupling strength. The difference combination resonance occurs when the
coupling between oscillators in the equation of motion is antisymmetric [see
Eqs. (\ref{x1}) and (\ref{x2})]. Such antisymmetric coupling is realized in
the Hamiltonian system if the two oscillators have the opposite sign of
masses. If coupling is symmetric (both masses are positive) the system has
the sum (rather than the difference) combination resonance.

We also find that generation of higher frequencies by means of the
difference combination resonance is possible in nonlinear materials with
positive refractive index but strong anomalous dispersion. In such a medium
the resonance can occur if at one of the generated frequencies the group and
phase velocities are opposite to each other. Anomalous dispersion is easier
to realize than the negative refractive index. Thus, highly dispersive
nonlinear optical systems are an attractive tool for applications in
generation of high frequency coherent radiation.

The sum combination resonance is a robust phenomenon which occurs in various
areas of physics. In contrast, the difference combination resonance is hard
to achieve because it requires special conditions, e.g., nonreciprocal
coupling. Recently such a resonance has been experimentally demonstrated in
nonreciprocally coupled resonant RLC circuits at radio frequencies %
\citep{Svidzinsky13}. It has also been proposed for generation of high
frequency coherent radiation in atomic medium by driving the atomic ensemble
with a much smaller frequency \citep{Svidzinsky13}. The proposed new kind of
light amplifier (called the QASER), contrary to a laser, does not need any
population of atoms in the excited state. The amplification mechanism of the
QASER is governed by the difference combination resonance which occurs when
the driving field frequency matches the frequency difference between two
normal modes of the coupled light atom system.

The present paper links the QASER amplification mechanism and parametric
processes that can occur in negative index nonlinear materials and strongly
dispersive media. The difference combination resonance in such systems opens
a perspective for development of new kind of light sources that, contrary to
conventional OPOs, emit coherent radiation of tunable wavelengths with the
frequency higher than the input frequency.

\begin{acknowledgments}
We thank Prof. Wolfgang Schleich for useful discussions. We gratefully
acknowledge support of the National Science Foundation Grants PHY-1241032
(INSPIRE CREATIV), PHY-1205868, and the Robert A. Welch Foundation (Awards
A-1261 and A-1547).
\end{acknowledgments}

\appendix

\section{Combination resonance in nonuniform medium}

Here we investigate conditions of the combination resonance in nonlinear
medium assuming that magnetic permeability $\mu (\mathbf{r})$ and dielectric
permittivity $\varepsilon (\mathbf{r})$ depend on coordinates. This allows
us to include nonuniform dielectric structures into consideration. For
simplicity we assume that dispersion is weak which allows us to do
calculations in the time domain. Maxwell's equations (\ref{me}) yield 
\begin{equation}
\text{curl}\left( \frac{1}{\mu }\text{curl}\mathbf{E}\right) +\frac{1}{%
c^{2}\varepsilon _{0}}\frac{\partial ^{2}\mathbf{D}}{\partial t^{2}}=0.
\label{c3a}
\end{equation}%
We write the total electric field as%
\begin{equation}
\mathbf{E}=\mathbf{E}_{d}(t,\mathbf{r})+\mathbf{E}_{1}(t,\mathbf{r})+\mathbf{%
E}_{2}(t,\mathbf{r}),
\end{equation}%
where%
\begin{equation}
\mathbf{E}_{d}(t,\mathbf{r})=\mathbf{A}_{d}(\mathbf{r})e^{i\nu _{d}t}+c.c.,
\end{equation}%
\begin{equation}
\mathbf{E}_{1}(t,\mathbf{r})=\mathbf{A}_{1}(\mathbf{r})e^{\mp i\omega
_{1}t+Gt},
\end{equation}%
\begin{equation}
\mathbf{E}_{2}(t,\mathbf{r})=\mathbf{A}_{2}(\mathbf{r})e^{i\omega _{2}t+Gt},
\end{equation}%
and $G$ is a small gain. We assume that the driving field frequency $\nu
_{d} $ satisfies the condition of the combination resonance

\begin{equation}
\nu _{d}=\omega _{2}\pm \omega _{1}.
\end{equation}%
Taking into account that nonlinear susceptibility $\chi $ is small and
disregarding the fast oscillating terms we obtain the following equations
for $\mathbf{A}_{1}(\mathbf{r})$ and $\mathbf{A}_{2}(\mathbf{r})$%
\begin{equation}
-\text{curl}\left( \frac{1}{\mu _{1}}\text{curl}\mathbf{A}_{1}\right) +\frac{%
\varepsilon _{1}\left( \omega _{1}\pm iG\right) ^{2}}{c^{2}}\mathbf{A}%
_{1}+\chi ^{\ast }\frac{\omega _{1}^{2}}{c^{2}}A_{d}^{\ast }\mathbf{A}_{2}=0,
\label{a1}
\end{equation}%
\begin{equation}
-\text{curl}\left( \frac{1}{\mu _{2}}\text{curl}\mathbf{A}_{2}\right) +\frac{%
\varepsilon _{2}\left( \omega _{2}-iG\right) ^{2}}{c^{2}}\mathbf{A}_{2}+\chi 
\frac{\omega _{2}^{2}}{c^{2}}A_{d}\mathbf{A}_{1}=0,  \label{a2}
\end{equation}%
where $\varepsilon _{1}$, $\mu _{1}$ and $\varepsilon _{2}$, $\mu _{2}$ are
the values of $\varepsilon $ and $\mu $ at the frequencies $\omega _{1}$ and 
$\omega _{2}$ respectively. Multiplying Eq. (\ref{a1}) by $A_{1}^{\ast }$
and Eq. (\ref{a2}) by $A_{2}^{\ast }$ yields%
\begin{equation}
-\frac{1}{\omega _{1}^{2}}\mathbf{A}_{1}^{\ast }\cdot \text{curl}\left( 
\frac{1}{\mu _{1}}\text{curl}\mathbf{A}_{1}\right) +\frac{\varepsilon _{1}(%
\mathbf{r})\left( \omega _{1}\pm iG\right) ^{2}}{\omega _{1}^{2}c^{2}}|%
\mathbf{A}_{1}|^{2}+\frac{\chi ^{\ast }}{c^{2}}A_{d}^{\ast }\mathbf{A}%
_{1}^{\ast }\mathbf{A}_{2}=0,  \label{a3}
\end{equation}%
\begin{equation}
-\frac{1}{\omega _{2}^{2}}\mathbf{A}_{2}\cdot \text{curl}\left( \frac{1}{\mu
_{2}}\text{curl}\mathbf{A}_{2}^{\ast }\right) +\frac{\varepsilon _{2}(%
\mathbf{r})\left( \omega _{2}+iG\right) ^{2}}{\omega _{2}^{2}c^{2}}|\mathbf{A%
}_{2}|^{2}+\frac{\chi ^{\ast }}{c^{2}}A_{d}^{\ast }\mathbf{A}_{1}^{\ast }%
\mathbf{A}_{2}=0.  \label{a4}
\end{equation}%
Subtracting these equations from each other, taking into account that 
\begin{equation}
\mathbf{A}^{\ast }\cdot \text{curl}\left( \frac{1}{\mu }\text{curl}\mathbf{A}%
\right) =\frac{1}{\mu }\left\vert \text{curl}\mathbf{A}\right\vert ^{2}+%
\text{div}\left( \frac{1}{\mu }\text{curl}\mathbf{A}\times \mathbf{A}^{\ast
}\right) ,
\end{equation}%
and integrating over space we find%
\begin{equation*}
\int d\mathbf{r}\left[ \frac{1}{\omega _{2}^{2}\mu _{2}}\left\vert \text{curl%
}\mathbf{A}_{2}\right\vert ^{2}-\frac{1}{\omega _{1}^{2}\mu _{1}}\left\vert 
\text{curl}\mathbf{A}_{1}\right\vert ^{2}\right. +
\end{equation*}%
\begin{equation}
\left. \frac{\varepsilon _{1}(\mathbf{r})\left( \omega _{1}\pm iG\right) ^{2}%
}{\omega _{1}^{2}c^{2}}|\mathbf{A}_{1}|^{2}-\frac{\varepsilon _{2}(\mathbf{r}%
)\left( \omega _{2}+iG\right) ^{2}}{\omega _{2}^{2}c^{2}}|\mathbf{A}_{2}|^{2}%
\right] =0.  \label{a0}
\end{equation}%
Taking imaginary part of Eq. (\ref{a0}), we then obtain%
\begin{equation}
\int d\mathbf{r}\left[ \frac{\varepsilon _{1}}{\omega _{1}}|\mathbf{A}%
_{1}|^{2}\mp \frac{\varepsilon _{2}}{\omega _{2}}|\mathbf{A}_{2}|^{2}\right]
=0.  \label{a5}
\end{equation}%
Here the upper (lower) sign corresponds to the sum (difference) combination
resonance.

On the other hand, taking real part of Eqs. (\ref{a3}) and (\ref{a4}) and
disregarding the small terms caused by nonlinearity we find after
integration over space%
\begin{equation}
\int d\mathbf{r}\varepsilon _{1}|\mathbf{A}_{1}|^{2}=\frac{c^{2}}{\omega
_{1}^{2}}\int d\mathbf{r}\frac{1}{\mu _{1}}\left\vert \text{curl}\mathbf{A}%
_{1}\right\vert ^{2},  \label{a6}
\end{equation}

\begin{equation}
\int d\mathbf{r}\varepsilon _{2}|\mathbf{A}_{2}|^{2}=\frac{c^{2}}{\omega
_{2}^{2}}\int d\mathbf{r}\frac{1}{\mu _{2}}\left\vert \text{curl}\mathbf{A}%
_{2}\right\vert ^{2}.  \label{a7}
\end{equation}

Eq. (\ref{a5}) implies that for the difference combination resonance (lower
sign \textquotedblleft $+$\textquotedblright ) the integrals $\int d\mathbf{r%
}\varepsilon _{1}|\mathbf{A}_{1}|^{2}$ and $\int d\mathbf{r}\varepsilon _{2}|%
\mathbf{A}_{2}|^{2}$ must have opposite signs. If, e.g., $\int d\mathbf{r}%
\varepsilon _{1}|\mathbf{A}_{1}|^{2}<0$ then Eq. (\ref{a6}) yields that $\mu
_{1}<0$ in some region of space. Thus, we obtain that the difference
combination resonance in nonlinear weakly dispersive materials is possible
only if for one of the generated frequencies both the magnetic permeability $%
\mu $ and dielectric permittivity $\varepsilon $ are negative in some
spatial volumes.

One should note that calculations presented here assume local (in time and
space) constitutive relations $\mathbf{D}=\varepsilon (\mathbf{r}%
)\varepsilon _{0}\mathbf{E}$ and $\mathbf{B}=\mu (\mathbf{r})\mu _{0}\mathbf{%
H}$. Thus, the results obtained are valid if one can disregard frequency and
spatial dispersion.

\end{document}